\documentclass[%
  aps,%
  prl,%
  twocolumn,%
  floatfix, %
  superscriptaddress,%
  longbibliography,
  nofootinbib,
]{revtex4-2}

\makeatletter
\def\doauthor#1#2#3{%
  \ignorespaces#1\unskip
  \begingroup
   #3
  \@if@empty{#2}{\endgroup{}{}}{\endgroup{\comma@space}{}\frontmatter@footnote{#2}}%
  \@listcomma
  \space \@listand
}%
\makeatother

\usepackage{amsmath,amssymb}
\usepackage{bbm} 
\usepackage{graphicx}
\usepackage{dblfloatfix} 
\usepackage{capt-of} 
\usepackage[caption=false]{subfig}
\usepackage{algpseudocode}

\newcounter{algorithm}
\renewcommand{\thealgorithm}{\arabic{algorithm}}
\newenvironment{algo}[1]{%
  \refstepcounter{algorithm}%
  \par\medskip\noindent\hrule\vspace{0.5ex}
  \noindent\textbf{Algorithm \thealgorithm.} #1\par\smallskip
}{%
  \vspace{0.5ex}\hrule\par\medskip
}
\usepackage{booktabs}
\usepackage{xcolor}
\usepackage{bm}
\usepackage[colorlinks=true,linkcolor=blue,citecolor=blue,urlcolor=blue]{hyperref}

\makeatletter
\let\revtex@section\section
\renewcommand\section{\@startsection{section}{1}{\z@}%
  {0.6ex \@plus .2ex \@minus .2ex}
  {-1em}
  {\itshape}}%
\makeatother

\AtBeginDocument{%
  \setlength{\abovedisplayskip}{2pt plus 1pt minus 1pt}%
  \setlength{\belowdisplayskip}{2pt plus 1pt minus 1pt}%
  \setlength{\abovedisplayshortskip}{0pt plus 1pt minus 1pt}%
  \setlength{\belowdisplayshortskip}{0pt plus 1pt minus 1pt}%
}

\newcommand{\chitop}{\chi_{\rm top}}

\begin{document}

\title{Sector-Resolved Flow Sampling for Topologically Frozen Lattice Gauge Theories}

\author{Ankur Singha}
\affiliation{Berlin Institute for the Foundations of Learning and Data (BIFOLD), Germany}
\affiliation{Technische Universit\"{at} Berlin, Germany}

\author{Jacob Kauffmann}%
\affiliation{Berlin Institute for the Foundations of Learning and Data (BIFOLD), Germany}
\affiliation{Technische Universit\"{at} Berlin, Germany}

\author{Karl Jansen}
\affiliation{Deutsches Elektronen-Synchrotron (DESY), Platanenallee 6,
15738 Zeuthen, Germany}
\affiliation{Computation-Based Science and Technology Research Center, The Cyprus Institute, Nicosia, Cyprus}

\author{Jacob Finkenrath}
\affiliation{Bergische Universit\"at Wuppertal, Germany}

\author{Vipul Arora}
\affiliation{Department of Electrical Engineering (ESAT), KU Leuven, Belgium}

\author{Shinichi Nakajima}

\affiliation{Berlin Institute for the Foundations of Learning and Data (BIFOLD), Germany}
\affiliation{Technische Universit\"{at} Berlin, Germany}
\affiliation{RIKEN Center for Advanced Intelligence Project (AIP), Japan}

\date{\today}

\begin{abstract}
Topological fluctuations are essential to nonperturbative gauge theories but become increasingly difficult to sample toward the continuum limit, where Markov chains can freeze in sectors of fixed topological charge. We introduce a generative sampler, a mixture of sector-resolved samplers (MSRS) that explicitly resolves these sectors and exploits a key advantage of generative models, the ability to directly evaluate the domain-restricted partition function and thereby determine the relative weights of disconnected sectors. We train a generative model in a reference topological sector combined with a bijective topological shift that deterministically maps its samples to other sectors.
We demonstrate the method in two-dimensional compact $U(1)$ lattice gauge theory, where it reproduces the topological-charge distribution and yields an unbiased susceptibility in a regime where hybrid Monte Carlo is frozen and overrelaxation gives inaccurate estimates. Our approach also outperforms existing flow-based samplers by orders of magnitude. These results demonstrate that explicit sector resolution provides a promising route to overcoming topological barriers in lattice gauge theory.
\end{abstract}

\maketitle

\section{Introduction}

Topology plays a central role in the nonperturbative dynamics of gauge theories. In quantum chromodynamics, fluctuations of the topological charge are intimately connected to the axial anomaly, the large mass of the $\eta'$ meson, and the dependence of the vacuum energy on the CP-violating angle $\theta$~\cite{Vicari2009,DelDebbio2002Theta}. First-principles investigations of such phenomena in lattice field theory rely heavily on numerical sampling, most commonly Markov-chain Monte Carlo (MCMC)~\cite{WilsonExact,PhysRevLett.62.361,duane1987hybrid}, to evaluate observables in high-dimensional configuration spaces. Reliable calculations require samplers to accurately capture fluctuations within each topological sector while correctly reproducing the relative statistical weights of the different sectors.
This requirement becomes increasingly difficult to satisfy toward the continuum limit~\cite{Alles1996,Schaefer2011,Gupta1992,DelDebbio2004}, as transitions between configurations with different topological charges become progressively suppressed, causing MCMC simulations to remain trapped in nearly fixed-topology sectors~\cite{Luscher2011,DelDebbio2004,Alexandrou2020,Bonanno2024wHMC,Hasenbusch2017}. 
Such topological freezing is a particularly important manifestation of a broader sampling problem encountered in lattice field theories and statistical systems, where physically relevant regions of configuration space are separated by large action or free-energy barriers~\cite{Nicoli:2023qsl}.

A variety of approaches have been developed to alleviate topological freezing. Open boundary conditions~\cite{Luscher2011}, tempering in the boundary conditions~\cite{Hasenbusch2017,Bonanno2021}, and metadynamics or related biased-sampling schemes~\cite{Laio2016,Eichhorn2026eok,Eichhorn:2022NH} enhance communication between sectors by modifying the boundary conditions or sampling dynamics. Other methods based on \emph{overrelaxation}, such as winding hybrid Monte Carlo (wHMC)~\cite{Bonanno2024wHMC} and Instanton Update~\cite{Eichhorn:2022NH}, 
introduce dedicated global updates designed to directly facilitate changes in topological charge. Although these methods can substantially improve the exploration of topological sectors, they ultimately rely on facilitating transitions 
between different sectors, whose occurence becomes increasingly suppressed at large action parameter value or toward the continuum limit for QCD.


Generative models offer a conceptually different approach by generating 
independent configurations without relying on 
Markov-chain evolution~\cite{Albergo2019,Kanwar2020,Nicoli2021,Caselle2022,Singha2023,gerdes2023learning,Wang2024Diffusion,Finkenrath2022,kanaujia2024advnf,Vaitl_2022,Finkenrath2024Review,Singha2026slc}. Gauge-equivariant normalizing flows~\cite{Kanwar2020, Boyda2021,Favoni2020reg, Abbott2023, Singha:2023xxq} have demonstrated the potential of direct sampling for topological observables in two-dimensional compact $U(1)$ gauge theory~\cite{Kanwar2020,Singha2023U1,Gerdes:2024rjk}, while subsequent work has introduced multiscale architectures to improve expressivity and scalability of generative models~\cite{Abbott2023,singha2025multilevel,bauer2025super,Singha2025RiGCS,Hasenfratz2026azk,Hasenfratz2026mae,BIALAS2022108502,Abbott:2024kfc}. Related generative strategies include stochastic normalizing flows inspired by non-equilibrium updates~\cite{Caselle2022,Bulgarelli2025,caselle2024sampling,Caselle:2024ent}, diffusion models connected to stochastic quantization~\cite{Wang2024Diffusion,aarts2026,Zhu:2025pmw,Kanwar:2025wuc,vega2025group,Tan:2026ibs}, gauge-covariant neural architectures~\cite{Tomiya2021}, and conditional flows transferred across couplings~\cite{Singha2023,gerdes2023learning,Faraz:2023xdi}.
Independent sample generation by generative models, however, does not by itself resolve the topological sampling problem, as the model
must capture both the fluctuations within each topological sector and the relative probabilities of multiple sectors. This requirement is particularly challenging under mode-seeking training objectives such as the reverse Kullback--Leibler (KL) divergence, which can underrepresent relevant sectors  or make the sampler concentrate predominantly within a single sector. More expressive 
model architectures can improve sampling performance, but their gains become increasingly limited as the separation between topological sectors grows, indicating that architectural refinement alone might not be sufficient to resolve the underlying sector-weighting problem~\cite{Kanwar2020,Abbott2023,Bonanno2026Flow}.


Here, we introduce a novel generative framework for sampling topological sector called a mixture of sector-resolved samplers (MSRS) by training sector-specific models and explicitly reconstructing their relative statistical weights. The key ingredient is the ability of the generative models to directly estimate the partition function~\cite{Nicoli2021} of each sector, thereby allowing the relative weights of the sectors to be determined.
In this work we train a generative model within a single reference sector and apply a transformation that is bijective over most of the probability mass to map the generated configurations to other sectors.  With these sector-specific samplers, we can estimate the sector-restricted partition function of each sector, yielding a mixture of samplers with accurate statistical weights. 


We demonstrate the framework in two-dimensional compact U(1) lattice gauge theory with the Wilson action. This theory exhibits severe topological freezing at large $\beta$~\cite{Gupta1992,Sinclair1992,Bonanno2024wHMC}, while exact finite-volume expressions for the topological-sector probabilities and susceptibility are available~\cite{WilsonExact}, providing an ideal test bed for validating the proposed framework. 
We benchmark our method against standard hybrid Monte Carlo (HMC), HMC with overrelaxation---including winding HMC~\cite{Bonanno2024wHMC} and HMC with IU~\cite{SMIT1987485,eichhorn2021comparison},
a conventional gauge-equivariant flow sampler~\cite{Kanwar2020}, and its more scalable multiscale extension~\cite{Abbott2023}. 
Across couplings extending deep into the frozen regime, our sampler reproduces the exact topological-charge distribution and yields unbiased susceptibility estimates while maintaining high sampling efficiency. In contrast, competing methods either become trapped and produce biased estimates by failing to reproduce the relative sector weights, or suffer orders-of-magnitude losses in efficiency.
The construction MSRS does not rely on a particular generative architecture and can extend to the theory beyond the one studied here.

\section{Theory and topological freezing}
\label{sec:setup}

Consider a lattice gauge theory with gauge group $G$ defined on a periodic lattice. The fundamental degrees of freedom are link variables $U_\mu(x)\in G$, distributed according to $p[U]=Z^{-1}e^{-S[U]}$. The action, and hence the probability distribution, is invariant under the gauge transformation $U_\mu(x)\rightarrow g^\dagger(x)U_\mu(x)g(x+\hat\mu)$, with $g(x)\in G$. The elementary gauge-invariant variables are the
plaquettes
{%
\setlength{\abovedisplayskip}{4pt}%
\setlength{\belowdisplayskip}{4pt}%
\setlength{\abovedisplayshortskip}{4pt}%
\setlength{\belowdisplayshortskip}{4pt}%
\begin{equation}
  \Pi_{\mu\nu}(x)
  =
  U_\mu(x)U_\nu(x+\hat\mu)
  U_\mu^\dagger(x+\hat\nu)U_\nu^\dagger(x).
  \notag
\end{equation}
}%
 In theories with nontrivial topology, configuration space decomposes into sectors labeled by an integer charge $Q[U]$. The relative statistical weights of these sectors define the probability distribution $P(Q)$, whose variance per unit volume is the topological susceptibility
\begin{equation} \chi_{\rm top} = \frac{\langle Q^2\rangle-\langle Q\rangle^2}{V}.
\end{equation}
In 2D $U(1)$ lattice gauge theory , writing $U_\mu(x)=e^{i\theta_\mu(x)}$ and defining the plaquette angle $\varphi_j=\arg\Pi_j\in[-\pi,\pi)$, we can express the action and the topological charge as
\begin{equation}
  S[U]=-\beta\sum_j\cos\varphi_j,
  \qquad
  Q[U]=\frac{1}{2\pi}\sum_j\varphi_j\in\mathbb{Z}.
  \label{eq:wilson_action_charge}
\end{equation}

In the MCMC based approach such as HMC, changing $Q$ continuously requires at least one plaquette to cross the branch cut at $\varphi_j=\pm\pi$ through a sector boundary. For the Wilson action, this entails an action barrier \(\Delta S_{\rm barrier}\simeq 2\beta\), which grows linearly with \(\beta\). Such topology-changing trajectories are therefore increasingly suppressed at large \(\beta\), leading to topological freezing.

Crucially, the increasing barrier between sectors does not imply that sectors with $Q\neq 0$ become statistically irrelevant. At large $\beta$, the minimum-action configuration in sector $Q$ distributes the total flux $2\pi Q$ approximately uniformly over the lattice, with $\varphi_j\simeq 2\pi Q/V$ (see Supplemental Material), yielding (see Fig.~\ref{fig:pipeline})
\begin{equation}
\Delta S_{\mathrm{gap}} \equiv
  S_Q^{\rm min}-S_{Q=0}^{\rm min}
  \simeq
  \frac{2\pi^2\beta Q^2}{V}.
  \label{eq:sector_action}
\end{equation}
Thus, the equilibrium suppression of a sector is controlled by $\beta/V$, whereas the barrier to reaching it grows as $\beta$. Along a line constant physics with $\beta/V=const.$, sectors with different $Q$ retain $O(1)$ relative weights even at large $\beta$, where HMC transitions between them become prohibitively rare. Consequently, MCMC based approaches may sample one sector accurately yet fail to reproduce the charge distribution $P(Q)$, leading to unreliable estimates of $\chi_{\rm top}$ and other topology-sensitive observables.


\section{Proposed Method}
\label{sec:method}

\begin{figure}[tb]
  \centering
\includegraphics[width=\columnwidth]{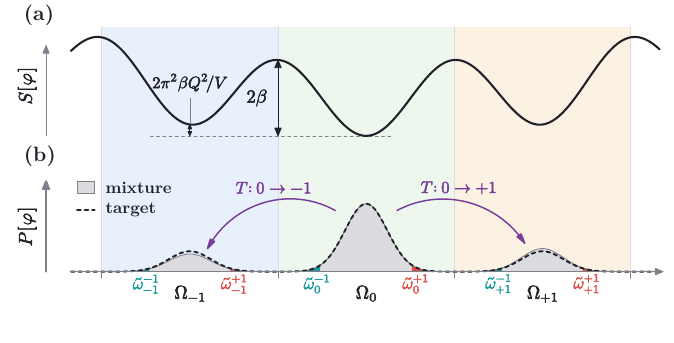}
  \caption{
  Action (top) and probability density (bottom) across different topological sectors.   In MSRS, the reference flow generates samples within $\Omega_0$, which are transferred to other sectors via $Q$-shift.  The mapping $T$ is bijective and volume-preserving, except in the action-suppressed regions $\tilde{\omega}_Q^{\pm1}$ (red and green).
  }
  \label{fig:pipeline}
\end{figure}

We work in the plaquette phase space $\Omega = \{\varphi \in [-\pi, \pi)^V; \sum_j\varphi_j=0\;(\mathrm{mod}\,2\pi) \}$, which is the $V$-dimensional torus restricted by the Bianchi identity and is in one-to-one correspondence with the link space through tree gauge fixing (see Appendix~\ref{app:plaquette}).%
\footnote{
 In our preliminary experiments, we found that 
plaquette-space sampling scales substantially better than 
direct link-space sampling. 
}
Throughout, $p[\varphi]$ and $r_{\psi}[\varphi]$ denote probability densities with respect to the intrinsic measure $d\mu [\varphi]$ on the $(V-1)$-dimensional Bianchi-constrained plaquette space $\Omega$.

Let $\Omega = \{\Omega_{Q}\}_{Q \in \mathbb{Z}}$ denote the decomposition of the plaquette space into topological sectors,
where 
$\Omega_{q} = \{\varphi \in \Omega; Q[\varphi] = q\}$.
Taking $\Omega_0$ as the reference sector, we train a normalizing flow $r_{\psi, 0}[\varphi]$ to generate exclusively within the $Q=0$ sector, by using the reverse-KL objective supplemented by a sector-localization penalty,
\begin{equation}
  \mathcal{L}(\psi)=\big\langle S[\varphi]+\log r_{\psi, 0}[\varphi]\big\rangle_{r_\psi}
  +\lambda_c\big\langle Q_{\mathrm{soft}}^2[\varphi]\big\rangle_{r_{\psi, 0}}
  \label{eq:rkl}
\end{equation}
Here, $Q_{\mathrm{soft}}$ is a differentiable surrogate for $Q$.
After training,
generated configurations $\varphi \sim r_{\psi, 0}$ such that $Q[\varphi] \ne 0$ are discarded, and $r_{\psi, 0}$ should be renormalized accordingly. Our approach avoids the need to train on all topological sectors simultaneously, as is typically done in existing generative approaches~\cite{Kanwar2020,Boyda2021,Singha2023U1}.




To transform configurations to other sectors, we
follow ~\cite{SMIT1987485,eichhorn2021comparison}
and introduce the following $Q$-shift operator $T_{Q_1 \to Q_2}: \tilde{\Omega}_{Q_1} \mapsto \tilde{\Omega}_{Q_2}$ such that
\begin{equation}
[T_{Q_1 \to Q_2}[\varphi]]_j= \textstyle \operatorname\!\varphi_j+\frac{2\pi \Delta Q}{V},
  \qquad j=1,\dots,V ,
  \label{eq:shift}
\end{equation}
where $\Delta Q = Q_2 - Q_1$ and
\begin{align}
\tilde{\Omega}_{\min(Q_1, Q_2)}^{|\Delta Q|}
&= \{\varphi \in \Omega_{\min(Q_1, Q_2)}; \max_j  \phi_j < \textstyle \pi - \frac{2\pi |\Delta Q|}{V}\},
\notag\\
\tilde{\Omega}_{\max(Q_1, Q_2)}^{-|\Delta Q|}
&= \{\varphi \in \Omega_{\max(Q_1, Q_2)}; \min_j  \phi_j \geq \textstyle - \pi + \frac{2\pi |\Delta Q|}{V}\}.
\notag
\end{align}
Crucially,
this mapping is bijective and volume preserving,
\begin{equation}
  T_{Q_1 \to Q_2}^{-1}=T_{Q_2 \to Q_1},\quad
  \big|\det \partial(T_{Q_1 \to Q_2} \varphi)/\partial \varphi\big|=1,
  \label{eq:Tq_props}
\end{equation}
and its domain and image cover most of the probability mass of the corresponding sectors, i.e., 
$P(\Omega_Q) \approx P(\tilde{\Omega}_Q^{|\Delta Q|})$, where $P(\Omega) \equiv \int_{\Omega} d \mu[\varphi] p[\varphi] $,
because any configuration in $\tilde{\omega}_Q^{\Delta Q} = \Omega_Q\backslash \tilde{\Omega}_Q^{\Delta Q}$ contains at least one plaquette close to the branch cut and is therefore strongly suppressed by the action, as illustrated in Figure~\ref{fig:pipeline}.

Accordingly,
ignoring the action-suppressed region in each sector, the $Q$-shift combined with the reference sampler $r_{\psi,0}$ provides a sector-specific sampler $r_{\psi, Q}
    =(T_{0 \to Q})_{\#} r_{\psi, 0}$ for any $Q \ne 0$ with the corresponding sampling density given by 
$
    r_{\psi, Q}[\varphi]
    =
    r_{\psi,0}(T^{-1}_{0 \to Q}[\varphi])$
 for $\varphi \in \tilde{\Omega}_Q^{-Q}$  (and $r_{\psi, Q}[\varphi] = 0$ for  $\varphi \notin \tilde{\Omega}_Q^{-Q}$).
Optionally, the reference sampler can be further specialized for each sector by fine-tuning $\psi$ by minimizing reverse-KL 
$\big\langle S[\varphi]+\log r_{\psi, Q}[\varphi]\big\rangle_{r_{\psi, Q}}$.

With the sector-specfic samplers $\{r_{\psi, Q}\}_{Q \in \mathbb{Z}}$, we apply the direct partition function estimator \cite{Nicoli2021} to the sector-restricted partition functions,
\begin{align}
    Z_Q= \textstyle \int_{\Omega_Q}d\mu(\varphi)\,e^{-S[\varphi]}
    \approx
   \frac{1}{N} \sum_{n=1}^N \frac{e^{-S[\varphi^{(n)}]}}{ r_{\psi, Q}[\varphi^{(n)}]}, \varphi^{(n)} \sim r_{\psi, Q},
   \notag
\end{align}
yielding an accurate estimate of the topological charge distribution
$P(Q) =  \textstyle \frac{Z_Q}{\sum_{q \in \mathbb{Z} }Z_q}$.
This completes the construction of a mixture of sector-specific samplers,
\begin{align}
    r_{\psi} [\varphi]
    &=\textstyle  \sum_{q} P(q) r_{\psi, q}[\varphi].
    \notag
\end{align}
In practice, sectors with large $|Q|$ and thus negligible probability mass can be excluded throughout the training process.
Further details of 
the training and sampling procedures can be found in Appendix~\ref{app:training_sampling}.

\section{Results}
\label{sec:results}

We first test the sampler along a line of constant physics (LCP), defined by
$\beta/L^2 = 0.094$, using six lattice sizes
$L\in{8,12,16,20,24,32}$
and the corresponding couplings
$\beta\in{6,13.5,24,37.5,54,96}$. 
This setup probes the approach to the continuum limit at fixed physical volume, where conventional algorithms become increasingly affected by topological freezing.

\onecolumngrid
\begin{center}
  \includegraphics[width=\textwidth]{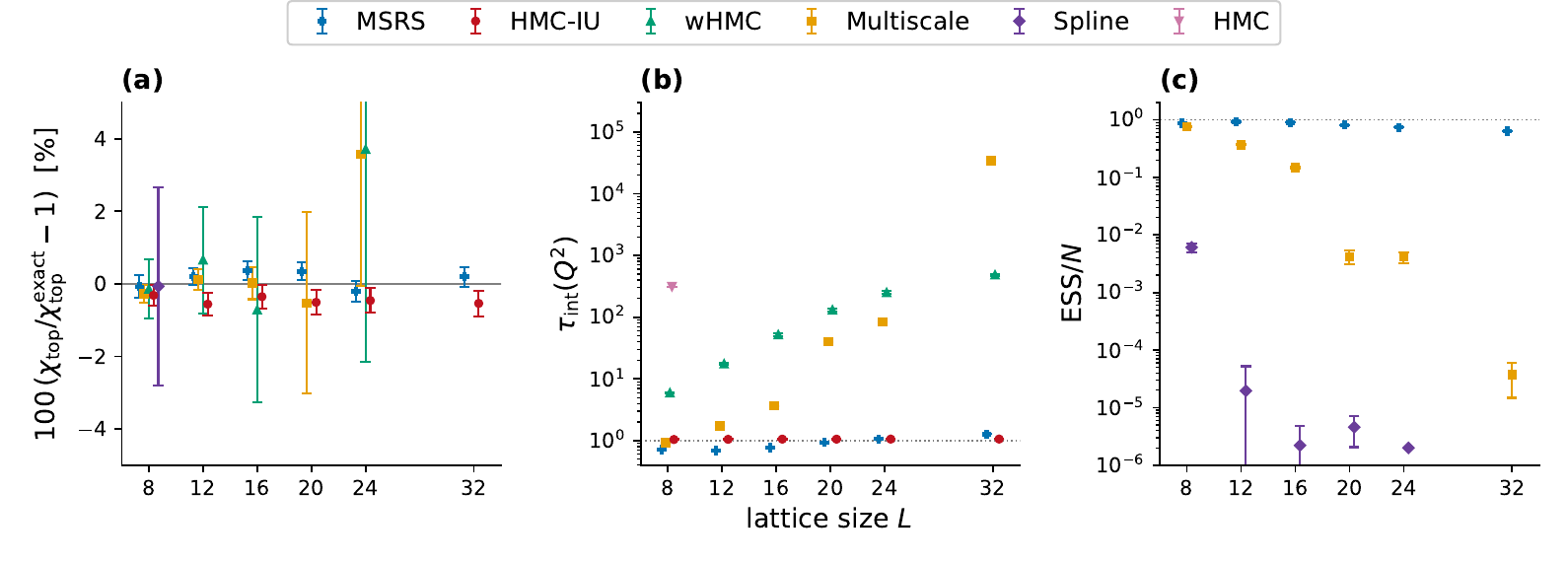}
  \captionof{figure}{Accuracy and efficiency of the topological-susceptibility estimate along
  the line of constant physics ($\beta/L^{2}=0.094$) as the continuum limit is
  approached.
  (a) Relative deviation of $\chi_{\rm top}$ from the exact Wilson value.
  (b) Integrated autocorrelation time $\tau_{\rm int}(Q^{2})$.
  (c) Effective sample size ESS$/N$.\label{fig:lcp_result}}
\end{center}
\twocolumngrid
Observables such as the topological susceptibilities are compared against the exact
finite-$\beta$, finite-$V$ Wilson value 
$\chitop^{\rm exact}$ obtained by
 character summation~\cite{Bonanno2024wHMC}. 
We benchmark our MSRS sampler against standard HMC, two overrelaxation schemes based on \(Q\)-shift updates—the Winding HMC (wHMC)~\cite{Bonanno2024wHMC} and HMC-IU (~\cite{eichhorn2021comparison}, where HMC-IU and MSRS uses the same Qshift; two flow-based baselines, a standard neural spline flow~\cite{Kanwar2020} and Multiscale flow following the coarse-to-fine construction of~\cite{Abbott2023} both trained with the reverse KL objective.


Figure~\ref{fig:lcp_result}(a) shows the relative deviation
of the topological susceptibility from its exact value.
Our MSRS, as well as HMC-IU~\cite{eichhorn2021comparison}, is compatible with the exact susceptibility at all lattice
sizes,
whereas the other baselines deteriorate rapidly as $L$ increases.
HMC-IU~\cite{eichhorn2021comparison} performs comparable to our approach, because it does not require the Markov chain to cross the high action barrier. However, it can suffer from low acceptance rates when the action difference \eqref{eq:sector_action} between sectors is large, while these low-probability sectors can still contribute significantly to susceptibility estimates.  This issue is expected to become more severe in higher-dimensional theories and larger gauge groups~\cite{Eichhorn:2022NH}. 
\begin{figure}[tbh]
  \centering
\includegraphics[width=\columnwidth]{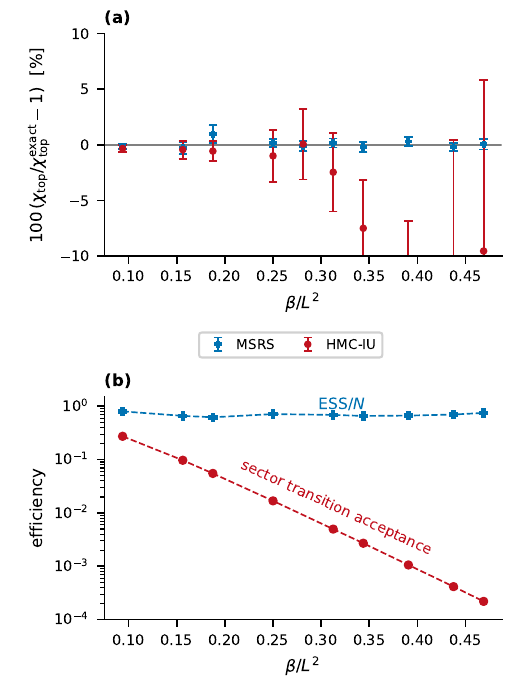}%
  \caption{  Deep frozen regime at fixed lattice size $L=8$, scanning $\beta/L^{2}$
  into large sector free energy $\Delta S_{\rm gap}$. (a) Relative deviation of $\chi_{\rm top}$ from the exact value, the
  IU develops a rapidly growing statistical uncertainty as the
  barrier increases, while MSRS's error stays small. (b) The IU's
  topological sector transition rate collapses by
  more than three orders of magnitude for large $\beta/L^2$, whereas MSRS's
  effective sample size ESS$/N$ remains flat.}
  \label{fig:frozen}
\end{figure}

Figures~\ref{fig:lcp_result}(b) and~\ref{fig:lcp_result}(c) show the autocorrelation times and effective sample sizes, respectively. MSRS achieves ESS values orders of magnitude larger than those of the competing generative approaches. While wHMC remains competitive in autocorrelation time on smaller lattices, its performance deteriorates with increasing lattice size. At the largest lattice size considered, all competing approaches become biased except HMC-IU, whose performance remains approximately flat across lattice sizes, similar to MSRS.

\begin{figure}[tbh]
  \centering
  \includegraphics[width=0.98\columnwidth]{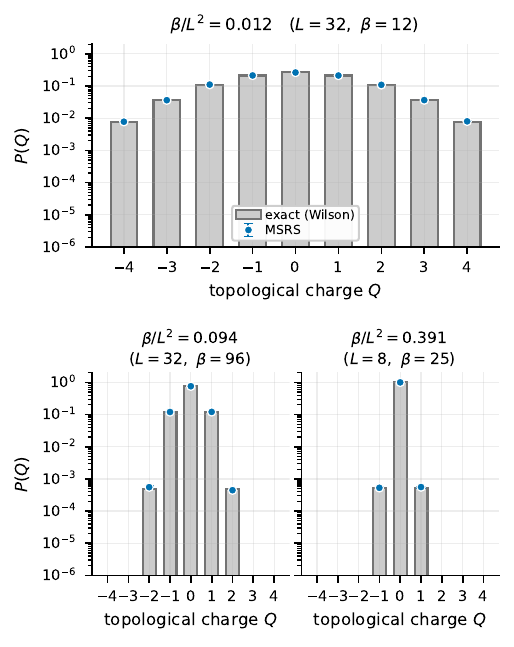}
  \caption{Comparison of the entire charge distribution $P(Q)$ with the exact Wilson result across three regimes, from broad to deeply frozen: $\beta/L^2=0.012$ and $0.094$ at $L=32$, and $0.39$ at $L=8$. The distribution narrows from support over $|Q|\le4$ to a regime in which $P(0)=0.999$ and the $|Q|=1$ sectors each carry probability $5\times10^{-4}$.
} 

  \label{fig:qdist}
\end{figure}

Figure~\ref{fig:frozen} tests the robustness of MSRS and the HMC-IU as the inter sector action gap between topological sectors increases. Fixing the lattice size at $L=8$, we evaluate both methods at progressively larger values of $\beta$. For $\beta/L^2\geq0.094$, the HMC-IU~\cite{eichhorn2021comparison} develops biased estimates or substantially increased variance, as shown in Fig.~\ref{fig:frozen}(a), consistent with the rapidly decreasing transition acceptance rates in Fig.~\ref{fig:frozen}(b). In contrast, MSRS remains insensitive to the growing inter-sector action gaps, with its ESS staying approximately constant across the couplings considered. This robustness follows from resolving the topological sectors explicitly: each mixture component targets an individual sector, while the relative sector weights are determined from estimates of the corresponding sector-restricted partition functions, including those of low-probability sectors.

To further validate the method, Fig.~\ref{fig:qdist} compares the full topological-charge distributions with the exact results across several physical volumes, including the extreme freezing regime, with agreement across all resolved sectors. These results demonstrate that explicit sector resolution provides a viable route to efficient sampling across topological sectors in lattice gauge theories. 



\section{Conclusion}
\label{sec:discussion}

We have introduced a sector-resolved generative sampler which combines a
normalizing flow trained within a single topological sector with a 
topology-changing transformation, forming a mixture of sector-specific generative samplers.
A key ingredient is the ability of normalizing flows to estimate the sector-resolving partition-function, highlighting the potential of machine learning to address the long-standing challenge of topological freezing.
We have demonstrated the performance of our samplers in different regimes, including the one with extreme topological freezing can persist even with efficient overrelaxation.

A natural next step is to generalize our approach to higher-dimensional theories and larger gauge groups, for which two major challenges are to be addressed.  First, for higher dimensional theories, we need to establish plaquette-based sampling that accounts for all constraints, such that the sampled plaquette space is in one-to-one correspondence with the gauge-fixed link space.  This is important because direct sampling in link space requires capturing long-range correlation extending across the entire volume in order to generate low-action plaquette configurations.
Plaqutte-space sampling, as employed in this work, can avoid this difficulty.
Second, $Q$-shift transformation needs to be generalized to larger gauge groups such as $SU(2)$ and $SU(3)$.

\section{Acknowledgments.}
This work was supported by the German Federal Ministry of Education and Research (BMBF) under grant BIFOLD25B and by the European Union’s Horizon Europe Marie Sk\l{}odowska-Curie Doctoral Networks programme through the AQTIVATE project (grant agreement No.~101072344). 
This work is also supported with funds from the Ministry of Science, Research, and Culture of the State of Brandenburg within the Centre for Quantum Technologies and Applications (CQTA).
This project received funding from the European Research Council (ERC) via the project ”LEEX” grant agreement 101170304. Funded by the European Union. Views and opinions expressed are however those of the author(s) only and do not necessarily reflect those of the European Union or the European Research Council Executive Agency (ERCEA). Neither the European Union nor the ERCEA can be held responsible for them.
We thank Stefan K\"uhn and Timo Eichhorn for helpful discussions and valuable comments on the manuscript.

\bibliographystyle{apsrev4-2}
\bibliography{refreneces}

\setcounter{secnumdepth}{1}
\makeatletter\let\section\revtex@section\makeatother

\setlength{\abovedisplayskip}{3pt plus 1pt minus 1pt}
\setlength{\belowdisplayskip}{3pt plus 1pt minus 1pt}
\setlength{\abovedisplayshortskip}{2pt plus 1pt minus 1pt}
\setlength{\belowdisplayshortskip}{2pt plus 1pt minus 1pt}

\clearpage
\phantomsection
\onecolumngrid
\begin{center}
  \Large Appendix
\end{center}
\vspace{0.8\baselineskip}

\twocolumngrid
\section{Tree gauge fixing: Plaquette Space sampling}
\label{app:plaquette}

We formulate the sampler in plaquette space and reconstruct the link configuration through a spanning-tree gauge fixing, yielding a one-to-one correspondence between plaquette and link variables (Fig.~\ref{fig:tree-gauge}).

To sample from the Bianchi-constrained plaquette space
$\Omega=\{\varphi\in[-\pi,\pi)^{V}:\sum_{j}\varphi_{j}=0\ (\mathrm{mod}\ 2\pi)\}$, we train a flow on a $(V-1)$-dimensional torus $\{\varphi\in[-\pi,\pi)^{V-1}\}$, and determine the final component $\varphi_{V}$ such that the Bianchi-identity is satisfied.
The corresponding link configurations are reconstructed as follows. We fix the gauge as well as the two Polyakov-loop angles, both of which leave the action invariant.  This allows us to fix $(V-1) + 2$ links: first, we set the links along the maximal spanning tree---shown in red in the right panel of Figure~\ref{fig:tree-gauge}---to zero.  We then set the two Polyakov-loop angles to zero, which fixes the links shown in purple to zero.
Next, following the order of the plaquettes numbered 1 to 15 in the figure, we determine the free links (green) from the corresponding plaqutte angles.  Since each plaquette contains only one such free link, these link values are determined without ambiguity.  This yields the complete link configuration, with the final plaquette angle automatically satisfying the Bianchi identity by construction.

\onecolumngrid
\begin{center}
  \vspace{-0.1\baselineskip}
  \includegraphics[width=0.99\textwidth]{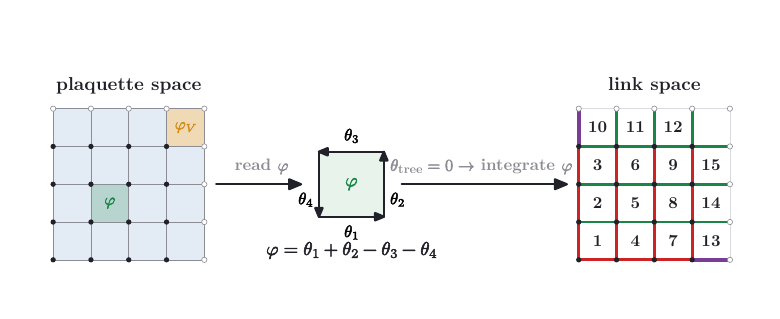}
  \vspace{-0.8\baselineskip}
  \captionof{figure}{ Plaquette-space sampling and its one-to-one mapping to the link space (after fixing the redundant degrees of freedom).  \textbf{Left: } A flow is trained on a $(V-1)$-dimensional torus, with the final component $\varphi_V$ determined by the Bianchi identity. \textbf{Middle: } Illustration of the relation between a plaquette and its constituent links.  \textbf{Right: } Corresponding reconstruction of the link configuration, with the gauge (red) and Polyakov-loop (purple) degrees of freedom fixed.
}
  \label{fig:tree-gauge}
\end{center}
\vspace{0pt}
\twocolumngrid
\section{Training and Sampling Details}
\label{app:training_sampling}
\paragraph{\textbf{Training (Alg.~\ref{alg:full})}.}The full method MSRS is summarized in Algorithm~\ref{alg:full}: we train a normalizing
flow $r_{\psi,0}$ localized in the reference sector $\Omega_0$
(Algorithm~\ref{alg:train}), map it to every sector by the unit-Jacobian
$Q$-shift, and estimate the sector partition functions to obtain the topological
charge distribution $P(Q)$. Observables are then drawn from the resulting mixture
(Algorithm~\ref{alg:sample}). Hyperparameters are given at the end of this section.

\begin{algo}{Training}\label{alg:full}
\begin{algorithmic}[1]
\State \textbf{Given:} action $S$, volume $V$, sector cutoff $K$, budget $M_0$
\State Train $r_{\psi,0}$ localized in $\Omega_0$ \Comment{Alg.~\ref{alg:train}}
\For{$Q=-K,\dots,K$}
  \State $r_{\psi,Q}(\varphi)=r_{\psi,0}\!\big(T^{-1}_{0\to Q}\varphi\big)$,\;
         $[T_{0\to Q}\varphi]_j=\varphi_j+2\pi Q/V$ \Comment{$|\det T_{0\to Q}|=1$, Eq.~\eqref{eq:shift}}
  \State draw $y_{1:M_0}\sim r_{\psi,0}$;\quad
         $\displaystyle Z_Q=\frac1{M_0}\sum_{i} \frac{e^{-S[T_{0\to Q}y_i]}}{r_{\psi,0}(y_i)}$
         \Comment{ using estimator~\cite{Nicoli2021}}
\EndFor
\State \Return sector samplers $\{r_{\psi,Q}\}_{|Q|\le K}$ and
       $P(Q)=Z_Q\big/\sum_{q}Z_q$
\end{algorithmic}
\end{algo}

\paragraph{\textbf{Reference sector Training (Alg.~\ref{alg:train})}.}
The sector-resolved estimator builds every sector sampler from a single reference
flow through the $Q$-shift, so training must produce a flow that lives entirely in
one topological sector; we therefore force the flow to collapse onto $\Omega_0$.

To drive the collapse we add to the reverse Kullback--Leibler objective
[Eq.~\eqref{eq:rkl}] a localization penalty $\lambda\,\langle Q_{\mathrm{soft}}^{2}\rangle$
built from the corner-plaquette surrogate
$Q_{\mathrm{soft}}=\varphi^{(\mathrm{corner})}/2\pi$. The tree gauge carries all
winding onto that plaquette: the Bianchi constraint fixes its angle to $-s$
($s=\sum_{j}\varphi_{j}$) and the integer charge is its winding,
$Q=\mathrm{round}(s/2\pi)=-\mathrm{round}(Q_{\mathrm{soft}})$, so penalizing
$Q_{\mathrm{soft}}^{2}$ localizes the flow at the sector centre $s=0$.

The penalty is reduced as
\begin{equation}
  \lambda_t=\lambda_{\min}+(\lambda-\lambda_{\min})\,e^{-t/\tau},
  \label{eq:lamschedule}
  \lambda_{\min}>0,
\end{equation}
The learning rate decays
geometrically and we retain the best post-warmup effective-sample-size
checkpoint. Finally the $Q\neq0$ leakage ($<0.04\%$ at every volume) is discarded and density is re-normalized.

The flow $r_{\psi,0}$ is a circular rational-quadratic neural spline flow ($36$
coupling layers, $8$ knots, hidden width $16$) with a uniform base, identical for
all $L$; only the training schedule and localization penalty vary with volume. In Fig.~\ref{fig:train_ess}, we show the learning curves for selected models along different LCPs.

\begin{algo}{Reference-sector flow training}\label{alg:train}
\begin{algorithmic}[1]
\State \textbf{Given:} $S$; steps $T$, warmup $W$; $\lambda,\lambda_{\min},\tau$;
       batch $B$; lr $\eta_0$, ratio $\rho$
\State initialize $\psi$;\; $\gamma=\rho^{1/(T-1)}$
\For{$t=1,\dots,T$}
  \State draw $\varphi_{1:B}\sim r_{\psi,0}$;\;
         $Q_{\mathrm{soft}}(\varphi_i)=\varphi_i^{(\mathrm{corner})}/2\pi$
  \State $\lambda_t=\lambda_{\min}+(\lambda-\lambda_{\min})e^{-t/\tau}$ if
         $\lambda_{\min}>0$, else $\lambda_t=\lambda\,[t\le W]$
  \State $\displaystyle\mathcal{L}=\frac1B\!\sum_i\!\big(S[\varphi_i]+\log r_{\psi,0}(\varphi_i)\big)
         +\lambda_t\frac1B\!\sum_i Q_{\mathrm{soft}}^2(\varphi_i)$
  \State $\psi\!\leftarrow\!\psi-\eta_t\nabla\mathcal{L}$;\;
         $\eta_{t+1}=\gamma\eta_t$;\; keep best-ESS $\psi$ if $t>W$
\EndFor
\State discard $Q\neq0$ leakage draws;\; \Return $\psi$
\end{algorithmic}
\end{algo}

\paragraph{\textbf{Sampling (Alg.~\ref{alg:sample})}.}
To estimate an observable we sample from the full mixture: for each draw we pick a
sector $Q\sim P(Q)$, generate a configuration from the reference flow $r_{\psi,0}$,
map it into that sector with the $Q$-shift, the density  for IU is given by $r_{\psi,Q}(\varphi)=r_{\psi,0}(T^{-1}_{0\to Q}\varphi)$. Sectors with $|Q|>K$ carry negligible mass, and we use
$K=4$. Bijective sector maps of this kind generalize naturally to theories with
higher symmetry groups and in higher dimensions~\cite{Eichhorn:2022NH}.

\begin{algo}{ Sampling}\label{alg:sample}
\begin{algorithmic}[1]
\State \textbf{Given:} $\{r_{\psi,Q}\}$, $P(Q)$, budget $N$, observable $\mathcal{O}$
\For{$n=1,\dots,N$}
  \State draw $Q_n\sim P$, $y_n\sim r_{\psi,0}$;\; $\varphi_n=T_{0\to Q_n}y_n$
  \State $w_n=e^{-S[\varphi_n]}\big/\big(P(Q_n)\,r_{\psi,0}(y_n)\big)$
\EndFor
\State \Return $\displaystyle\langle\mathcal{O}\rangle=\Big(\sum_n w_n\,\mathcal{O}[\varphi_n]\Big)\Big/\Big(\sum_n w_n\Big)$
\end{algorithmic}
\end{algo}

\paragraph{Hyperparameters.}
The architecture and optimizer (Adam, batch $64$, lr $10^{-3}\!\to\!10^{-5}$,
warmup $2\times10^3$, $2\times10^5$ steps) are shared across volumes; the
localization penalty is $\lambda=30$ with no persistent floor for $L\le24$, and is
raised to $\lambda=\lambda_{\min}=150$ only at $L=32$. For evaluation we use
$M_0=2\times10^5$ and $N=5\times10^5$ with $K=4$; errors are delete-block jackknife
and unbiasedness is judged as the median over five evaluation seeds.
\vspace{1\baselineskip}
\section{Extended results}
\label{app:extended_result}
\subsection{Higher moments and \texorpdfstring{$\theta$}{theta}-dependence}
\label{app:theta}
\begin{figure}[t]
  \centering
  \includegraphics[width=\columnwidth]{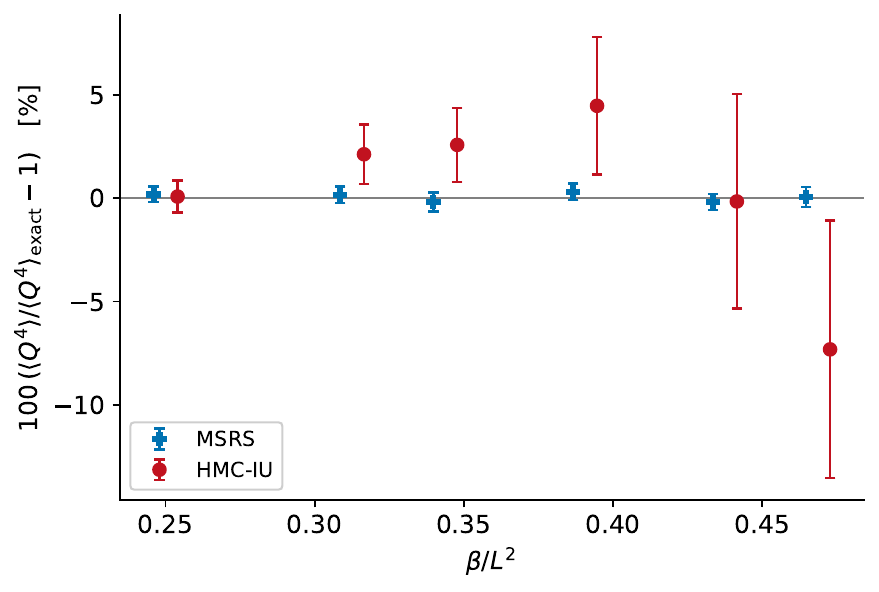}
  \caption{Fourth moment $\langle Q^{4}\rangle$ at $L=8$ for the instanton-update (IU) (with 3M samples) and
 MSRS (0.5M samples) against the exact finite-$\beta$ Wilson value}
  \label{fig:figEM}
\end{figure}

Resolving the relative weight of every sector gives access to higher moments of the
topological charge, and hence to the $\theta$-dependence beyond
$\chi_{\rm top}=\langle Q^{2}\rangle/V$. The fourth moment $\langle Q^{4}\rangle$ is the
first such quantity and is controlled by the rare high-$|Q|$ tail that a frozen Markov
chain fails to sample. As Figure~\ref{fig:figEM} shows, MSRS reproduces it to sub-percent,
while the instanton-update chain (IU) is noisier and, at the extreme frozen point, biased
low from never reaching the $|Q|\ge2$ sectors.

These moments fix the $\theta$-dependence of the vacuum free energy,
$f(\theta)-f(0)=\tfrac12\chi_{\rm top}\theta^{2}(1+b_{2}\theta^{2}+\dots)$, with
$b_{2}=-\langle Q^{4}\rangle_{c}/(12\langle Q^{2}\rangle)$ and
$\langle Q^{4}\rangle_{c}=\langle Q^{4}\rangle-3\langle Q^{2}\rangle^{2}$
($b_{2}=-1/12$ for a dilute gas). At these $L=8$ points the charge is confined to
$|Q|\le1$, MSRS resolves the higher moments at all, making the $b_{2}$ and the higher-order $\theta$-dependence accessible in frozen regimes where MCMC struggles.

 \vspace{-1.8\baselineskip}
\subsection{Further line of constant physics}
\label{app:multilcp}\vspace{-0.5\baselineskip}
Repeating the analysis of our MSRS at $\beta/L^{2}=0.156$ (Table~\ref{tab:extended}) keeps
$\chi_{\rm top}$ unbiased ($\lesssim1\%$), $\tau_{\rm int}(Q^{2})=\mathcal{O}(1)$, and
the effective sample size high across $L=8$--$24$; the other lines
($0.125,\,0.188$) are in the Supplemental Material.
\begin{table}[tbh]
\centering
\caption{$\beta/L^{2}=0.156$, $N=5\times10^{5}$. ESS
errors are delete-block jackknife; $\tau_{\rm int}$ errors are the Madras--Sokal estimate.}
\label{tab:extended}
\begin{ruledtabular}
\begin{tabular}{c ccc}
$L$ & ESS$/N$ & $\tau_{\rm int}(Q^{2})$ & $\delta\chi/\chi$ \\ \hline
$8$  & $0.666 \pm 0.007$ & $1.26 \pm 0.01$ & $-0.1\%$ \\
$12$ & $0.815 \pm 0.004$ & $0.89 \pm 0.01$ & $+0.2\%$ \\
$16$ & $0.838 \pm 0.003$ & $0.86 \pm 0.01$ & $-0.9\%$ \\
$20$ & $0.703 \pm 0.005$ & $1.12 \pm 0.01$ & $-0.8\%$ \\
$24$ & $0.512 \pm 0.006$ & $1.61 \pm 0.01$ & $+0.8\%$ \\
\end{tabular}
\end{ruledtabular}

\end{table}

\begin{center}
  \includegraphics[width=.49\textwidth]{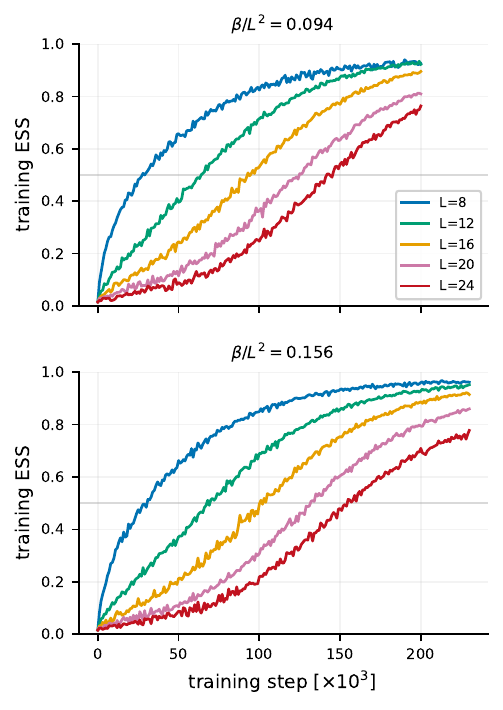}
  \captionof{figure}{Training dynamics: effective sample size (ESS) during training.}
  \label{fig:train_ess}
\end{center}
\twocolumngrid

\end{document}